\documentclass[11pt,letterpaper]{article}
\usepackage{emnlp2017}
\usepackage[utf8]{inputenc}
\usepackage{natbib}
\usepackage{url}
\usepackage{graphicx}
\usepackage{caption}
\usepackage{amsthm}
\usepackage{amssymb}
\usepackage{amsmath}
\usepackage{subcaption}
\usepackage{cleveref}
\usepackage{tabularx}
\usepackage{bbm}
\usepackage{comment}
\usepackage{todonotes}
\usepackage{floatrow}
\DeclareFloatFont{tiny}{\footnotesize}
\emnlpfinalcopy

\title{Risk Estimation of SARS-CoV-2 Transmission from Bluetooth Low Energy Measurements}
\author{Felix Sattler$^\dag$, Jackie Ma$^\dag$, Patrick Wagner$^{\ddag}$,\\ \bf David Neumann$^\dag$, Markus Wenzel$^\dag$, Ralf Sch\"{a}fer$^\dag$,\\ \bf Wojciech Samek$^{\dag*}$, Klaus-Robert M\"{u}ller$^{\ddag\S\sharp*}$, and Thomas Wiegand$^{\dag\ddag*}$\\[+6px] $^\dag$Fraunhofer Heinrich Hertz Institute, 10587 Berlin, Germany\\
$^\ddag$Technische Universit\"at Berlin, 10587 Berlin, Germany\\
$^\S$Department of Brain and Cognitive Engineering, Korea University, Seoul, Korea\\
$^\sharp$Max Planck Institute for Informatics, 66123 Saarbr\"ucken, Germany\\
{\tt\footnotesize \{wojciech.samek, thomas.wiegand\}@hhi.fraunhofer.de,} {\footnotesize\tt klaus-robert.mueller@tu-berlin.de}}
\date{}

\begin{document}
\maketitle
\begin{abstract}
Digital contact tracing approaches based on Bluetooth low energy (BLE) have the potential to efficiently contain and delay outbreaks of infectious diseases such as the ongoing SARS-CoV-2 pandemic. In this work we propose a novel machine learning based approach to reliably detect subjects that have spent enough time in close proximity to be at risk of being infected. Our study is an important proof of concept that will aid the battery of epidemiological policies aiming to slow down the rapid spread of COVID-19.
\end{abstract}

Contact tracing is an effective instrument to contain and delay outbreaks of infectious diseases such as the ongoing SARS-CoV-2 pandemic. Individuals that have been in contact with an infected person are identified, asked to remain in quarantine and are being tested. However, manually following contact histories is labor-intensive, slow and incomplete, as chance encounters, e.g. in the public transport, can not be fully reconstructed. The emergence of digital solutions, which automatically reconstruct the duration and proximity of encounters, is highly promising to enhance established manual procedures with speed, efficiency, precision and full coverage of relevant contact history. Ultimately, such  proximity tracing technologies have the potential to ``reduce transmission enough to achieve $R_0 < 1$ and sustained epidemic suppression, stopping the virus from spreading further''  ~\citep[]{Ferretti2020}.

Various concepts for proximity tracing have been proposed in the past \citep[e.g.][]{salathe2010high, yoneki2011fluphone, EBOLAPP, TraceTogether, chen2018next}. Recently, the \emph{Pan-European Privacy-Preserving Proximity Tracing} (see \citep[]{pepppt}) and \emph{Decentralized Privacy Preserving Proximity Tracing} (see \citep[]{dp3t}) initiatives were launched, both promising to enable proximity tracing in compliance with the European general data protection regulation (GDPR) \cite{voigt2017eu}. Since a large percentage of the world's population carries smartphones, these approaches make use of the BLE technology. Contact advertisements regularly emitted from these devices are used to assess the proximity of encounters. For effectively containing the current SARS-CoV-2 pandemic, it is necessary to reliably translate the BLE signal strength measurements into risk estimates of infection transmission.
In this letter, we propose a novel approach for this conversion task.

\begin{figure*}[th!]
\centering
\includegraphics[width=1\textwidth]{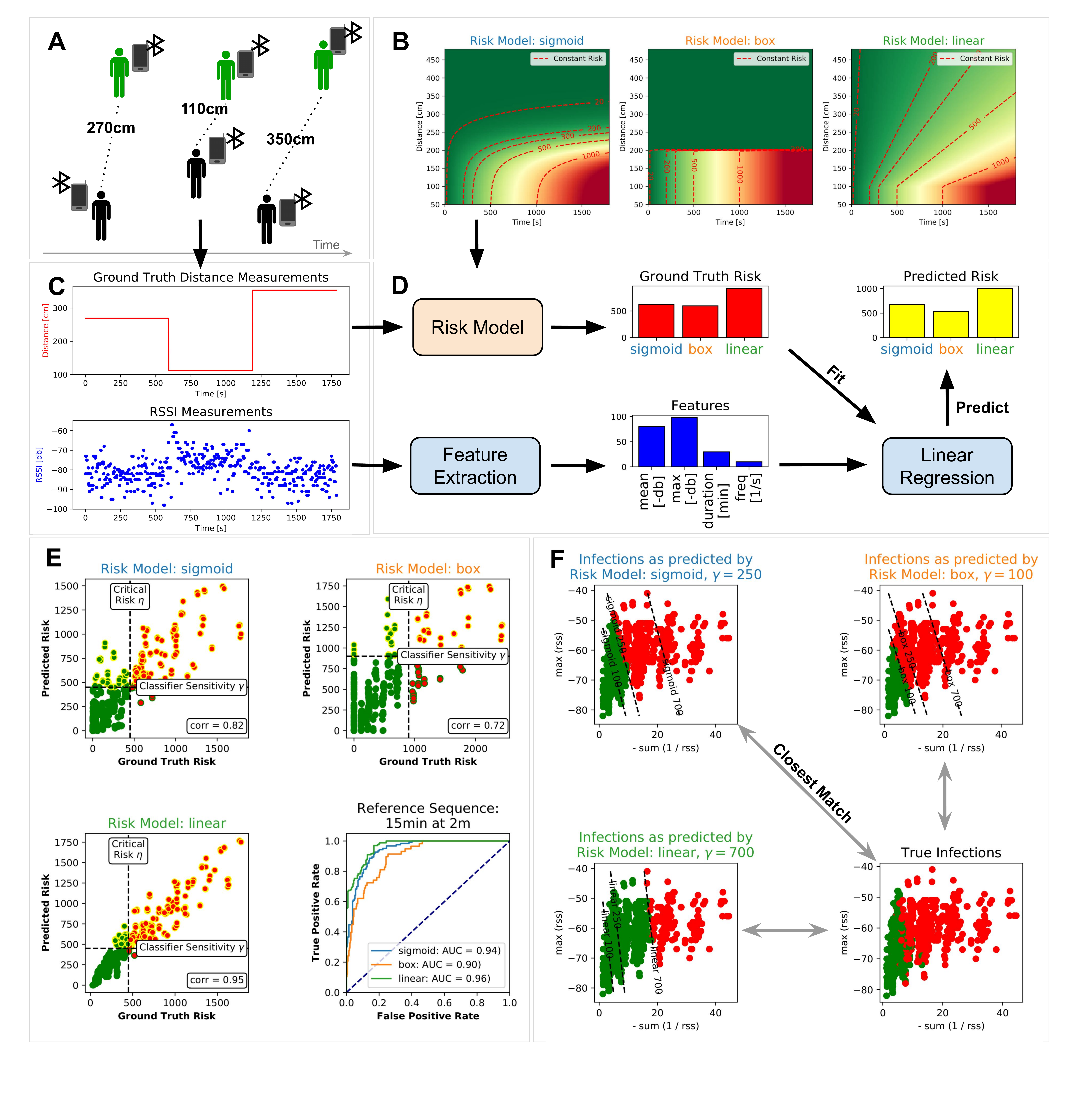}
\caption{Overview of the proximity tracing concept and results. \textbf{A}: Typical infection scenario in a public space (e.g. a supermarket), where close contact ($<$ 2m) between an infected and a contact person is established over a long enough period of time.
\textbf{B}: An epidemiological model translates a time series of contact distances into infectiousness scores, which are then used to label the encounters in the training dataset. 
\textbf{C}: Example of a raw RSS time series of the BLE signal, as well a corresponding contact distances.
\textbf{D}: We train a linear regression model to predict the infectiousness scores obtained from a given risk model. The linear regression receives as input a list of features, which were derived from the raw rss data.
\textbf{E}: The predictions of the linear regression model correlate strongly with the ground truth risk (up to 0.95 for the linear risk model). For a fixed critical risk threshold $\eta$ the approach achieves high true positive rates with very few false classifications. \textbf{F}: To this day only little is known about spreading behaviour of SARS-Cov-2. In this work, we calibrated our epidemiological models according to the latest recommendations of epidemiologists \cite{ecdc}. After large-scale deployment of proximity tracing technologies, it will be possible to compare the predicted infection events with the actually measured ones. This may help to refine epidemiological models.}\label{fig:results}
\end{figure*}

Figure \ref{fig:results}A illustrates a typical infection scenario, which is difficult to manage with manual contact tracing procedures. Here, an infected person enters a public place (e.g. a supermarket) and spends an extended amount of time in close proximity ($<$ 2m) to the contact person. Both factors, namely the contact distance and the contact duration, influence the risk for the contact person of being infected.

Proximity tracing technologies allow to reconstruct such high risk encounters between the infected and contact person, once the former has been tested positive. The infected person is recording anonymous IDs of contact persons within certain critical distance range. These anonymous proximity histories are encrypted and remain on the phone of the infected person at all times. Only if tested positive and upon agreement, the proximity history is analyzed and contact persons with a high risk of being infected can be alerted anonymously. In addition, health authorities can be involved to handle these high risk cases by standard procedures (e.g., test and quarantine the contact persons). 

\vspace*{0.2cm}
\noindent
\textbf{Methods}:
\noindent

To make this approach practically applicable, i.e., to avoid that every short time or distant encounter raises an alarm, it is crucial to reliably estimate the risk of infection transmission from the BLE signal strength measurements. In this letter we propose to perform this conversion in the following manner:
\begin{enumerate}
    \item {\bf Define an epidemiological model to convert proximity time series to infection risk scores}.
    The models $E$ displayed in Figure \ref{fig:results}B implement different non-linear functions to translate time series of proximity values into {\it infection risk scores}. For infections transmitted via the droplet route, one usually assumes that the infection risk decreases as the distance $d_t$ between people increases; with some critical distance from which on the risk of being infected becomes vanishingly low \cite{xie2007far}.
    \item {\bf Use the epidemiological model for data labeling}. The chosen epidemiological model is then used to label the data needed to train the ML-based infection risk predictor. For that, one integrates the marginal infection risk within the critical distance over the contact duration $T$ to obtain an infection risk score
$$
I = \sum_{t=1}^T E(d_{t}).
$$
An encounter between two individuals is considered as ``high risk'' if the value of $I$ exceeds a predefined critical risk threshold $\eta$. This threshold can either be set locally, i.e., for each encounter\footnote{For COVID-19 it is assumed that a physical proximity between two people of less than 2 meters over a time period of 900 seconds (15 minutes) results in a high risk of being infected \cite{ecdc}. When setting $\eta$ locally, one would use these parameters to determine if an encounter is labelled as ``high risk'' or not.}, or globally based on the basic reproduction rate\footnote{The $R_0$ value tells us the expected number of cases directly generated by one case. A globally set critical risk $\eta$, would label the data such that the number of ``high risk'' encounters would match the expected number of cases.} $R_0$. 
    
    \item {\bf Train a machine learning (ML) model to estimate the infection risk}. We train a linear regression model to predict the infection risk score from the measured received signal strength (RSS) time series of the BLE signal. For simplicity, we do not provide the raw RSS time series to the ML model, but compute features (sum, mean, max etc.) on it and provide this aggregated information to the model. By thresholding $\gamma$, the output of the classifier, we can trade-off the obtained true and false positive rates. Figure \ref{fig:results}D illustrates the entire training and evaluation pipeline, including ground truth risk estimation, feature extraction and training of the linear regression model.   
\end{enumerate}

Figure \ref{fig:results}C displays the time series of raw RSS values from the BLE signal, which the smartphone of the infected person receives from the smartphone of the contact person. Although the values are noisy, it is possible to reliably decide whether or not the infection risk $I$ exceeds a certain threshold, as shown in our real-world experiments performed with 48 participants (see supplementary materials for the details on the experimental setup). 

Figure \ref{fig:results}E, compares the ground truth risk, as computed from the time series of ground truth distances, with the predicted risk, estimated from the Bluetooth signal strength data, for 392 contact episodes from a holdout validation set. As we can see, our machine learning based approach, is able to achieve correlation numbers of up to 0.95 for the linear infection risk model. We compute the critical risk threshold $\eta$ by inserting the reference sequence $d^{ref}$, with
\begin{equation}
\label{eq:seq_ref}
    d_t^{ref}\equiv 200cm\ \mathrm{and}\;\; T^{ref}=900s
\end{equation} into the different risk models. By varying the classifier sensitivity $\gamma$, we can trade-off the number of correct and false alarms. The resulting receiver operating characteristic (ROC) curve of the real-world experiment displayed in Figure \ref{fig:results}E shows that high true positive rates can be achieved with relatively few false classifications. Note that these ROC curves depend on the data labeling procedure, i.e., the epidemiological model and the threshold $\eta$. Here we used the assumed parameters for COVID-19, namely distance $<2$ m and exposure time $>15$ min \cite{ecdc}. We provide mean and maximum RSS value as well as the number of received Bluetooth beacons as features to the linear regression model; results with other features derived from the RSS time series can be found in the supplementary materials.
The AUC (area under the ROC curve) value of the predictor is found to be larger than 0.9 for all investigated epidemiological models. For the linear model AUCs of up to 0.96 were obtained. The prediction task becomes slightly more difficult for the \texttt{box} and \texttt{sigmoid} models, which assign only negligible risk to encounters above a certain distance. The repetition of this analysis on data recorded on another day led to very similar performance results, demonstrating the reliability of the proposed approach (see supplementary materials).

Figure \ref{fig:results}F displays RSS sequence data\footnote{Every RSS sequence is represented as a dot. Displayed are only two features of every RSS sequence, the maximum and the sum of the negative inverse RSS values.} along with the classification decisions of linear classifiers, which were trained to match the predictions of three different epidemiological models. These classification decisions also depend on the corresponding thresholds $\eta$. 
Clearly, once proximity tracing technologies will be rolled out for the broad population, the true infection events will be observed (given data donations and consent of all users involved). Comparing the observed infections with the predicted ones\footnote{Of course one could directly compare the risk scores derived from the epidemiological model with the observed infection events. However, to obtain the scores from the epidemiological model, one need proximity data which are unavailable in practice. The proposed approach is scalable, because it directly estimates the risk from RSS values.} will provide evidence for the true (and currently unknown) epidemiological modeling assumptions.

In this letter we have proposed a novel approach to reliably detect subjects that have spent enough time in close proximity to be at risk of being infected. Thus our study is an important proof of concept that will aid the battery of epidemiological policies aiming to slow down the rapid spread of COVID-19. Note that while we have assumed the standard modeling of viral spread with the currently agreed on parameters (distance $<2$ m and exposure time $>15$ min, see \cite{ecdc}), it may in fact be conceivable that these parameters are not chosen conservatively enough in the light of recent results on contagious droplet spreading across larger distances rsp.\ in aerosols (see e.g.~\cite{10.1001/jama.2020.4756}) and moreover the improved binding affinity of SARS-CoV-2 \cite{wrapp2020cryo}. Clearly, once proximity tracing technologies will be rolled out for the broad population, then data transmission events will become available that will provide evidence for the true epidemiological modeling assumptions. With that we could find out whether the current risk assessment is conservative enough or whether indeed social distancing would need to be increased further. 

\vspace*{0.2cm}
\noindent
\textbf{Data Availability}:
\noindent
The data that support the findings of this study are available from the corresponding author upon request.

\vspace*{0.2cm}
\noindent{\bf Acknowledgement.} This work was supported by the German Ministry for Education and Research as BIFOLD - Berlin Institute for the Foundations of Learning and Data (ref.\ 01IS18025A and ref.\ 01IS18037A). KRM also received support from the Institute for Information \& Communications Technology Promotion and funded by the Korea government (MSIT) (No.\ 2017-0-00451, No.\ 2017-0-01779). Correspondence to WS, KRM, TW.

\bibliographystyle{plainnat}
\bibliography{references}

\begin{thebibliography}{13}
\providecommand{\natexlab}[1]{#1}
\providecommand{\url}[1]{\texttt{#1}}
\expandafter\ifx\csname urlstyle\endcsname\relax
  \providecommand{\doi}[1]{doi: #1}\else
  \providecommand{\doi}{doi: \begingroup \urlstyle{rm}\Url}\fi

\bibitem[Bourouiba(2020)]{10.1001/jama.2020.4756}
Lydia Bourouiba.
\newblock {Turbulent Gas Clouds and Respiratory Pathogen Emissions: Potential
  Implications for Reducing Transmission of COVID-19}.
\newblock \emph{JAMA}, 03 2020.
\newblock ISSN 0098-7484.
\newblock \doi{10.1001/jama.2020.4756}.
\newblock URL \url{https://doi.org/10.1001/jama.2020.4756}.

\bibitem[Chen et~al.(2018)Chen, Yang, Pei, and Liu]{chen2018next}
Hechang Chen, Bo~Yang, Hongbin Pei, and Jiming Liu.
\newblock Next generation technology for epidemic prevention and control:
  Data-driven contact tracking.
\newblock \emph{IEEE Access}, 7:\penalty0 2633--2642, 2018.

\bibitem[DP-3T()]{dp3t}
DP-3T.
\newblock \url{https://github.com/DP-3T/documents}.

\bibitem[{European Centre for Disease Prevention and Control}(2020)]{ecdc}
{European Centre for Disease Prevention and Control}.
\newblock Contact tracing: public health management of persons, including
  healthcare workers, having had contact with covid-19 cases in the european
  union -- second update, 2020.

\bibitem[Ferretti et~al.(2020)Ferretti, Wymant, Kendall, Zhao, Nurtay,
  Abeler-Dörner, Parker, Bonsall, and Fraser]{Ferretti2020}
Luca Ferretti, Chris Wymant, Michelle Kendall, Lele Zhao, Anel Nurtay, Lucie
  Abeler-Dörner, Michael Parker, David~G Bonsall, and Christophe Fraser.
\newblock Quantifying {SARS-CoV-2} transmission suggests epidemic control with
  digital contact tracing.
\newblock \emph{Science}, 2020.
\newblock \doi{TBA}.
\newblock URL \url{TBA}.

\bibitem[{Freunde~Liberias, e.~V.}(2018)]{EBOLAPP}
{Freunde~Liberias, e.~V.}
\newblock {EBOLAPP}.
\newblock \url{https://www.ebolapp.org/}, 2018.

\bibitem[PEPP-PT()]{pepppt}
PEPP-PT.
\newblock \url{ https://www.pepp-pt.org}.

\bibitem[Salath{\'e} et~al.(2010)Salath{\'e}, Kazandjieva, Lee, Levis, Feldman,
  and Jones]{salathe2010high}
Marcel Salath{\'e}, Maria Kazandjieva, Jung~Woo Lee, Philip Levis, Marcus~W
  Feldman, and James~H Jones.
\newblock A high-resolution human contact network for infectious disease
  transmission.
\newblock \emph{Proceedings of the National Academy of Sciences}, 107\penalty0
  (51):\penalty0 22020--22025, 2010.

\bibitem[{Singapore Government Technology Agency} and {Ministry of
  Health}(2020)]{TraceTogether}
{Singapore Government Technology Agency} and {Ministry of Health}.
\newblock {TraceTogether}.
\newblock \url{https://www.tracetogether.gov.sg/}, 2020.

\bibitem[Voigt and Von~dem Bussche(2017)]{voigt2017eu}
Paul Voigt and Axel Von~dem Bussche.
\newblock The eu general data protection regulation (gdpr).
\newblock \emph{A Practical Guide, 1st Ed., Cham: Springer International
  Publishing}, 2017.

\bibitem[Wrapp et~al.(2020)Wrapp, Wang, Corbett, Goldsmith, Hsieh, Abiona,
  Graham, and McLellan]{wrapp2020cryo}
Daniel Wrapp, Nianshuang Wang, Kizzmekia~S Corbett, Jory~A Goldsmith, Ching-Lin
  Hsieh, Olubukola Abiona, Barney~S Graham, and Jason~S McLellan.
\newblock Cryo-em structure of the 2019-ncov spike in the prefusion
  conformation.
\newblock \emph{Science}, 367\penalty0 (6483):\penalty0 1260--1263, 2020.

\bibitem[Xie et~al.(2007)Xie, Li, Chwang, Ho, and Seto]{xie2007far}
X~Xie, Y~Li, AT~Chwang, PL~Ho, and WH~Seto.
\newblock How far droplets can move in indoor environments--revisiting the
  wells evaporation-falling curve.
\newblock \emph{Indoor air}, 17\penalty0 (3):\penalty0 211--225, 2007.

\bibitem[Yoneki(2011)]{yoneki2011fluphone}
Eiko Yoneki.
\newblock Fluphone study: Virtual disease spread using haggle.
\newblock In \emph{Proceedings of the 6th ACM Workshop on Challenged Networks},
  pages 65--66, 2011.

\end{thebibliography}

\onecolumn
\appendix
\section{Supplementary Material}
\subsection{Epidemiological Models}
In our experiments we use three different epidemiological models to convert the proximity values into infectiousness scores
\begin{equation}
    E_{linear}(d) = 
    \begin{cases}
    1 & \text{ if } d[cm] < 100\\
     \frac{100}{d[cm]} & \text{ else}
    \end{cases}
\end{equation}

\begin{equation}
    E_{box}(d) = 
    \begin{cases}
    1 & \text{ if }d[cm]\leq 200\\
    0 & \text{ else}
    \end{cases}
\end{equation}

\begin{equation}
    E_{sigmoid}(d) = \left (1 + \exp \left ( \frac{d[cm]-200}{30} \right ) \right )^{-1},
\end{equation}
where $d$ is the contact distance measured in cm.
All three models are monotonically decreasing functions of the distance and the infectiousness score decreases with increasing distance. 

The main use of epidemiological models in our experiments is to generate ground truth labels for our data, which consists of a time series of RSS values and corresponding distances (the latter is not available in real settings). To generate the labels, we integrate the infectiousness scores over the contact time according to the equation
\begin{equation}
\label{eq:1}
    I(d_1,..,d_T)  = \sum_{t=1}^T E(d_t).
\end{equation}

\subsubsection{Local and Global Risk Thresholds}
For every epidemiological model $E$ there exists a {\it reference}, from which on no infection is expected. For instance, for COVID-19 it is assumed that a physical proximity between two people of less than 2 meters over a time period of 900 seconds (15 minutes) results in a high risk of being infected \cite{ecdc}. Inserting the reference sequence $d^{ref}$, with
\begin{equation}
    d_t^{ref}\equiv200cm\ \mathrm{and}\;\; T^{ref}=900s
\end{equation}
into equation (\ref{eq:1}) results in a {\it local threshold}
\begin{equation}
    \eta = I(d^{ref}) = \sum_{t=1}^{T^{ref}} E(d_t^{ref}) 
\end{equation}

By selecting the epidemiological model and the infectiousness threshold we can determine, which time series of distance measurements should be considered dangerous and which should not:
\begin{equation}
    \textnormal{HighRisk}(d_1,..,d_T) = \begin{cases}
    True & \text{if } I(d_1,..,d_T)>\eta\\
    False & \text{if } I(d_1,..,d_T)\leq \eta
    \end{cases}
\end{equation}

An alternative approach is to label the data with a {\it global threshold}. For that we need to have an estimate of the expected number of newly infected contact persons $N_{new}$ from $N_{inf}$ previously infected persons. This number can be computed with the basic reproduction number $R_0$ as
\begin{equation}
N_{new} = R_0N_{inf}
\end{equation}
One can then chose $\eta$ in a way so that the number of high risk encounters matches the expected number of new infections, i.e.,
\begin{equation}
\sum_{d\in D}\mathbbm{1}_{\textnormal{HighRisk}(d)} = N_{new}
\end{equation}
where $D$ is the total number of recorded proximity histories.
\subsection{Infection Risk Estimation as a Regression Problem}
Given an epidemiological model and the true distances we can label encounters into ``high risk'' and ``low risk''. Since the true distances are not available in real settings, we aim to train a machine learning model to predict these labels from the raw\footnote{For practical reasons we resampled the RSS values to 1Hz.} RSS measurements of the BLE signal.
To simplify the learning task, we extract features from the RSS data and provide them as input to the ML algorithm. In particular, we tested the following three feature sets:
\begin{enumerate}
    \item \textbf{sum}: total sum of received RSS values resulting in one-dimensional features
    \item \textbf{dur\_max\_mean}: duration, maximum and mean of received RSS values resulting in three-dimensional features.
    \item \textbf{freq}: amplitudes of first $30$ frequencies of received RSS values resulting in $30$-dimensional features.
\end{enumerate}

We input these features into a linear regression model in order to obtain a predicted ``risk'' score:
\begin{equation}
    I^{predict}(RSS_1,..,RSS_T) = \langle A, f(RSS_1,..,RSS_T)\rangle + b
\end{equation}
The input to the linear regression thus comprises a vector of parameters $A$, a bias term $b$ and a vector of extracted features $f(RSS_1,..,RSS_T)$. The resulting predicted risk score is then compared to a threshold, which can be set to $\eta$. If the predicted risk exceeds the threshold the encounter which resulted in the sequence of RSS measurements is considered ``high risk``.

\subsection{Real-World Experiment}
\subsubsection{Experimental Setup}
A measurement campaign was performed to test and validate the proposed infection risk estimation model. This section describes the setup of the experiment. 

The measurements on the 1st of April and the 7th of April were performed using 48 Samsung A40 smartphones of the same type that were carried by 48 protected soldiers, respectively. Tests were carried out at five different locations within the Julius Leber barracks in Berlin. There were three rooms within a conference center and two outdoor locations, with ten subjects each. All test subjects were equipped with face masks so that there was no risk of infection.

\begin{figure}[h!]
     \centering
     \includegraphics[width=.7\textwidth]{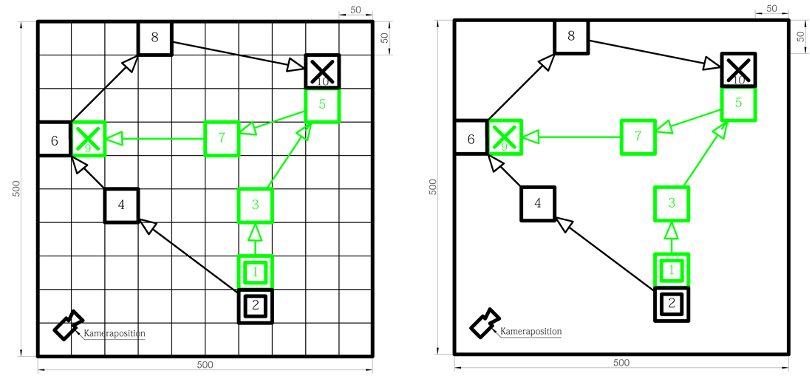}
     \caption{Test pattern on the floor of the five test areas (left with grid, right without grid).}
     \label{fig:test_pattern}
\end{figure}

The floor of the test areas was marked (Fig.\ \ref{fig:test_pattern}). These markings consisted of a 5 m x 5 m grid with lines spaced 50 cm apart. From the starting point (box within a box) to the ending point (multiplication sign), the test subjects had to walk through markings and stay on each marker for a predetermined amount of time (2, 4, 6, or 10 min). The markings are numbered on the green path from 1 to 9 and on the black path from 2 to 10 (Fig.\ \ref{fig:test_pattern}, right). Two cameras were installed at each location to video record the test so that the exact locations of the test subjects could be checked after the test. 
The test was carried out in four runs. During the runs, the test subjects were instructed not to move too much, to hold the positions of the mobile phones relatively stable, and to stand within the square. 

\subsubsection{Data and Preprocessing}
RSS data was collected via a prototype of the PEPP-PT App. The RSS data - recorded at a random and potentially varying frequency between 0.1 Hz and 10 Hz - was re-sampled to 1Hz. Ground truth distance data was derived from the predefined movement pattern on the grid. The labeling was additionally verified with the help of video footage that was taken at the test area. For every pair of soldiers we collected multiple data points, where one data point comprised of two aligned sequences:

\begin{itemize}
    \item A time series of distances $d_t, t=1,..,T$ (from which the ground truth risk can be derived).
    \item A time series of BLE RSS values $RSS_t, t=1,..,T$, recorded by mobile phones held by the soldiers.
\end{itemize}

\subsubsection{Training and Testing Data}
For training and testing, the time series data was separated into two folds according to the room in the test area in which the data was collected. Data collected in rooms 1 and 2 (indoor) and room 4 (outdoor) was combined in the training set. Data collected in rooms 3 (indoor) and 5 (outdoor) was combined in the validation set. In previous tests multiple combinations of indoor and outdoor rooms were tested to investigate possible covariate shift between indoor and outdoor scenarios. No significant effects could be detected, therefore the aforementioned mixed split was used.

\subsubsection{Results}
We trained a machine learning model to predict the ground truth risk, by only using features extracted from the RSS time series data $RSS_1,..,RSS_T$. Since the labels are not balanced (i.e.\ there are more negative than positive events), we use area under the ROC (receiver operating characteristics) curve (AUC) metric to evaluate the performance of our model. The AUC metric is a measure for how well the data can be separated using our classifier. An AUC value of $0.5$ indicates no predictive power and $1.0$ indicates perfect predictive power.

\begin{figure}[!ht]
    \centering
    \includegraphics[width=\textwidth]{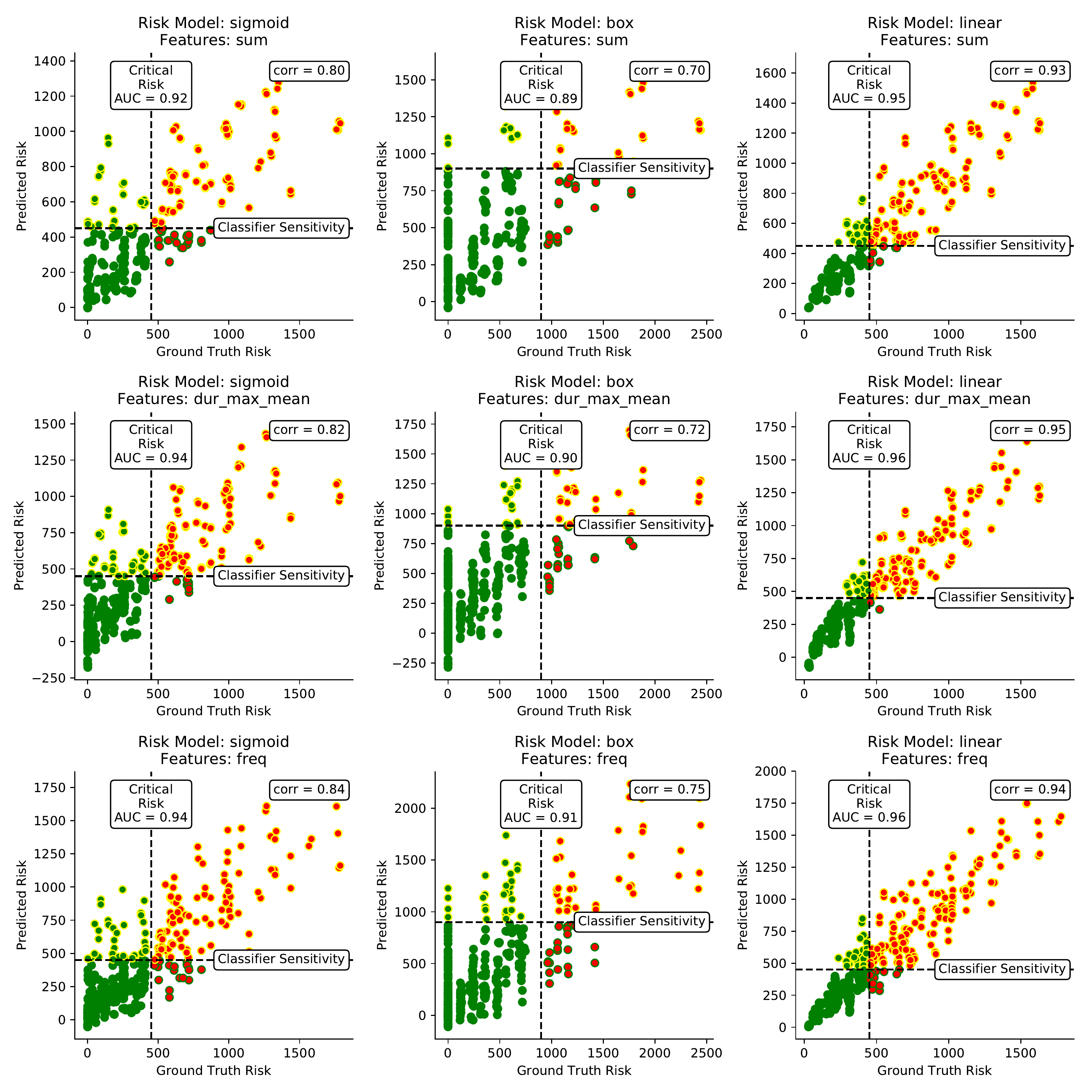}
    \caption{Ground truth risk vs predicted risk for different epidemiological risk models and combinations of features supplied to our machine learning model.}
    \label{fig:auc}
\end{figure}

The obtained results are presented in Fig. \ref{fig:auc}. The columns correspond to different epidemiological models, namely (\texttt{linear}, \texttt{box}, \texttt{sigmoid}), whereas the rows represent different combinations of features which we feed into the linear regression. Given the critical risk threshold derived by applying the respective risk model to the reference sequence \eqref{eq:seq_ref}, we display the achieved AUC for every combination of risk model and feature combination.

An encounter between two individuals is labeled as ``high risk'' if the value of $I$ exceeds a predefined critical risk threshold $\eta$. This threshold can either be set locally, i.e., for each encounter
, or globally based on the basic reproduction rate
$R_0$. 

\subsubsection{Follow-up Study}
\begin{figure}[h!]
    \centering
    \includegraphics[width=\textwidth]{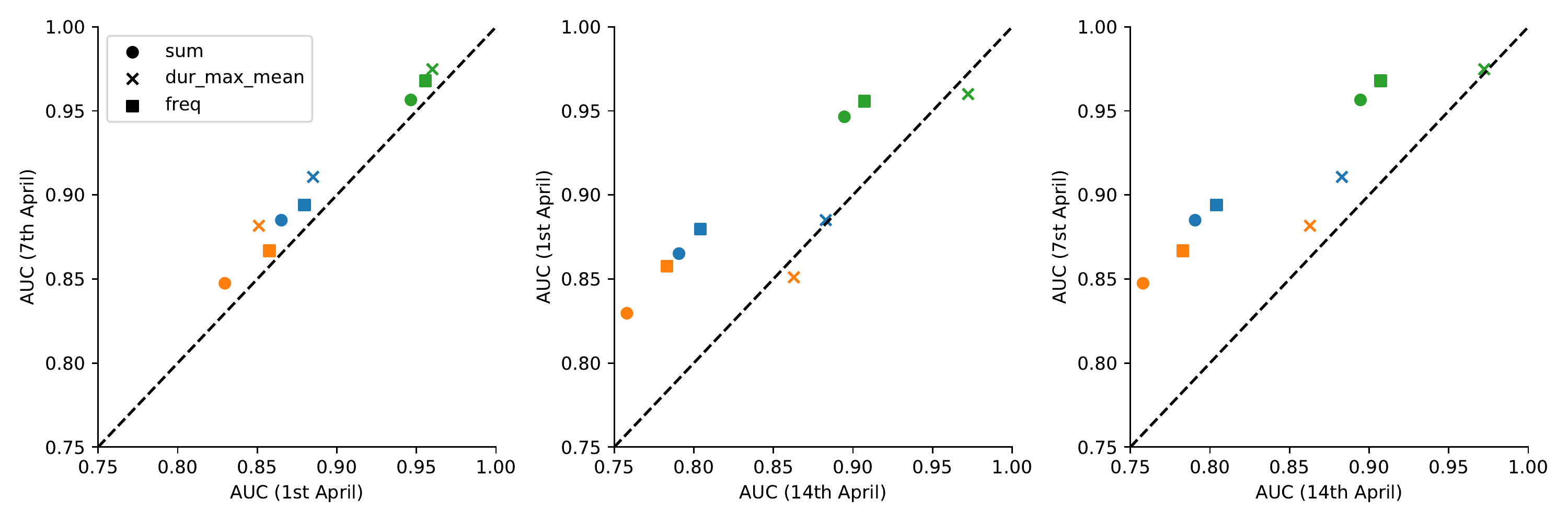}
    \caption{Comparison of the results on data recorded on three different days. Every marker corresponds to a epidemiological model ((\texttt{linear} - green, \texttt{box} - orange, \texttt{sigmoid} - blue) and a set of features (\texttt{sum}, \texttt{dur\_max\_mean}, \texttt{freq}). Only the feature set \texttt{dur\_max\_mean} is robust to the changes in testing environment that occurred during the third measurement campaign on April 14th.}
    \label{fig:additional}
\end{figure}

In order to evaluate the reliability of our results, we tested the model on data recorded with the same experimental setup, but on a different dates (7th April 2020 and 14th April 2020). In the experiments conducted during the 14th of April, participants were using different smart phone models and the phone holding positions were varied ("hand", "ear", "pocket"). Figure \ref{fig:additional} compares the AUC values of the two measurement campaigns for the three epidemiological models (\texttt{linear}, \texttt{box}, \texttt{sigmoid}) and three sets of features (\texttt{sum}, \texttt{dur\_max\_mean}, \texttt{freq}). As can be seen, the performance of the proposed infection risk estimation method is comparable for the experiments conducted on the 1st of April and the 7th of April. For the experiments conducted on the 14th of April however the feature set \texttt{dur\_max\_mean} distinctively outperforms all other tested feature combinations. Evidently this combination of features is able to approximate the ground truth risk in a more robust way than the other investigated feature combinations.

\end{document}